# Collective Intelligence as Infrastructure for Reducing Broad Global Catastrophic Risks

Vicky Chuqiao Yang [1]*, Anders Sandberg [2]

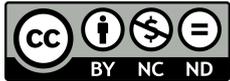

**Funding:** V.C.Y. was partially supported by NSF Grant 2117564

**Conflict of Interest Statement:** The authors report no conflict of interest

**Informed Consent Statement:** No human subjects were involved in this study

**Acknowledgments:** The authors thank Rory Greig for helpful discussions and feedback on an earlier version of the manuscript.

**Author Contributions:** V.C.Y. led the design of the research and the writing of the manuscript, while A.S significantly contributed to both efforts.

**Abstract:** Academic and philanthropic communities have grown increasingly concerned with global catastrophic risks (GCRs), including artificial intelligence safety, pandemics, biosecurity, and nuclear war. Outcomes of many, if not all, risk situations hinge on the performance of human groups, such as whether governments or scientific communities can work effectively. We propose to think about these issues as Collective Intelligence (CI) problems—of how to process distributed information effectively. CI is a transdisciplinary research area, whose application involves human and animal groups, markets, robotic swarms, collections of neurons, and other distributed systems. In this article, we argue that improving CI in human groups can improve general resilience against a wide variety of risks. We summarize findings from the CI literature on conditions that improve human group performance, and discuss ways existing CI findings may be applied to GCR mitigation. We also suggest several directions for future research at the exciting intersection of these two emerging fields.

[1] Assistant Professor, MIT Sloan School of Management, Massachusetts Institute of Technology, 100 Main Street, Cambridge MA, USA; vcyang@mit.edu

[2] Senior Research Fellow, Future of Humanity Institute, University of Oxford, Oxford, UK, anders.sandberg@philosophy.ox.ac.uk

* Correspondence: vcyang@mit.edu

> *We've got to be as clear-headed about human beings as possible, because we are still each other's only hope.  — James Baldwin*

## 1. Global Catastrophic Risks

Recent years have witnessed increasing concerns for humanity's survival and prosperity in the long-term future. Scholars are increasingly concerned about global catastrophic events. Examples of such events include nuclear wars, climate catastrophes leading to large-scale failures in agriculture, pandemics, and powerful artificial intelligence (AI) not aligned with human values (see Ord, 2020 for a summary). Other scholars are also concerned with future technological innovations that enable a single individual to have great destructive power, such as DIY biohacking tools that can kill millions (Bostrom, 2019). Scholars studying these potentially catastrophic events and how to mitigate them have formed a transdisciplinary field called global catastrophic risks (GCR), and a subset focused on events that can lead to human extinction is referred to as existential risks. An associated, more practice-focused field, effective altruism, investigates how an individual can do the most good to help others. These areas are not only of concern to academics but have also attracted considerable attention and action in philanthropic efforts.

Discussion in GCR has been centered around which risks are likely to occur and how to circumvent each one. A critical component crosscuts many, if not all, risks but remains less explored—GCR reduction hinges on the better collective performance of human groups. For example, mitigating global pandemic risks includes coordinating individual lifestyle changes, the scientific community's research efforts, and nation-states' policies. We consider factors helping to mitigate many risk scenarios "infrastructures" of the system. Prior research investigated several other such factors, including several risks having the common denominator of disrupting agriculture (Denkenberger et al., 2021); the pace of regulatory innovation being far slower than that of technological innovation (Marchant, 2011); and dangerous technology being accessible to a small number of actors (Bostrom, 2019). The vulnerability of having a weak infrastructure, such as in collective decision-making and collective action, reduces the general stability of the system. Such a system can be compared to a tightrope walker—many forces can lead them to fall, be it a gust of wind or a shake of the body. Compared to focusing on which forces will tip the tightrope walker over and prevent these forces from happening, it is more useful to identify tools that can help improve the general stability of the system. For example, adding a balancing pole (Fig 1), which increases the toques needed that lead to a fall and expands the walker's ability to adjust their center of gravity through small pole movements. Similarly, in GCR mitigation, it is essential to consider building general tools, such as the capability for collective decision-making, that improve a system's ability to respond to change. This view is echoed in a perspective piece for collective behavior to be considered a "crisis discipline" (Bak-Coleman et al., 2021). The idea of a system's stability to any perturbation also goes beyond metaphors---it has been rigorously developed in the mathematical theory of dynamical systems, and there has been promising progress (albeit with limitations) in inferring the closeness of a system to its tipping point from time-series data (Pananos et al., 2017; Bury et al., 2022).

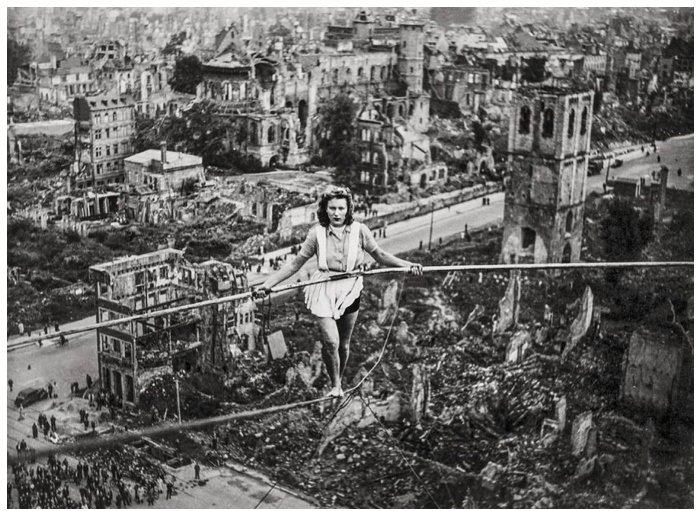

**Figure 1.** *Tightrope walker Margret Zimmermann over Köln in 1946 (Heukeshoven, 1946). The balance pole used by tightrope walkers serves as a device for improving the ability to adaptively balance and adjust to perturbations. The role of Collective Intelligence in Global Catastrophic Risk mitigation is similar to that of the balance pole to a tightrope walker—but performed by a multitude of actors.*

While most GCR mitigation efforts center on resolving technological challenges, a significant minority of the GCR community is concerned with human collective decision and action (Liu et al., 2018) and typically frame these issues as coordination problems that can be approached in a game-theoretic framework such as in prisoner's dilemma. While informative in many applications, it can be too narrow and restricting to think about these complex phenomena of collective decisions and action exclusively as problems of coordination. Firstly, this framework typically considers a small number or type of agents with a fixed set of explicit rules. While many hard problems we face are among a large number or type of agents, and the mechanisms of interaction may be implicit or change over time. This distinction is at the core of some most important human endeavors---humans created systems for unrelated individuals to cooperate at a large scale and in a wide range of ways (Melis & Semmann, 2010). This is unique among animals. For example, chimpanzees and other mammals can cooperate in flexible ways but in smaller groups; social insects, such as bees and ants, can work in large numbers but with more rigid roles. Second, besides the consideration of group size and flexible ways of interaction, another issue with viewing the human collective through the lens of coordination is that it tends to dwell on the negative outcomes of poor coordination, like the tragedy of the commons. These are important to avoid; however, making it the sole focus risks neglecting the upsides of the human collective, where the group is more capable than the sum of the individuals within it. The focus on avoiding bad coordination may lead to neglecting how to achieve that. Instead of thinking about these human-collective issues as an issue of coordination, we would advocate for thinking about them more broadly as an issue of Collective Intelligence, one of how to process distributed information effectively in order to reach common goals.

## 2. Collective Intelligence

Broadly speaking, Collective Intelligence (CI) is concerned with a group's ability to perform a wide variety of tasks or solve a wide variety of problems (Woolley et al., 2010). Since what are relevant problems are subjective to the observer, collective intelligence is referred to by some as groups of individuals acting collectively in ways that *seem* intelligent (Malone et al., 2009). This ability is typically associated with group synergy, that the group outperforms the capability of its individual members (Kurvers et al., 2015).

Examples of CI include democratic elections, prediction markets, and juries. It also goes beyond human groups to include animal groups, such as in a flock of birds deciding the direction for migration; robotic swarms, such as designing rules for individual robots such that the collective can perform certain tasks; and neurons, such as the brain making coherent sense of the world while each neuron responds to different, and sometimes conflicting stimuli (see Fig. 2 for illustrations of these examples). In all these diverse applications, a shared problem is when information is distributed across the individuals in the system, how to process distributed information effectively and aggregate it together. This distributed information processing can take a wide variety of forms spanning a wide range of complexity. Some examples of more explicit aggregation mechanisms include majority vote in political elections, and quorum voting in African buffalos when deciding on new grazing locations (one signals preference by standing up and facing the favored direction) (Prins, 1996). The more implicit mechanisms involve the schooling of fish, where each fish's preferred direction of movement interacts with that of its neighbors to generate movements of the school. As a result, the school can collectively sense the temperature gradient, while no individual fish is able to sense such information (Puckett et al., 2018). Another example of more implicit aggregation is prediction markets, where individuals bet on the likelihood of an outcome, and the information on the odds is embedded in the prices. Researchers have come together to study these phenomena, crosscutting application domains, and formed a transdisciplinary field around CI. We refer to this field as transdisciplinary instead of interdisciplinary, because it studies a shared phenomenon that manifests in many disciplines, spanning computer science, neuroscience, robotics, animal behavior, and many social sciences. It transcends the disciplinary boundaries, while interdisciplinary often refers to combining two fields to discover new things in either one.

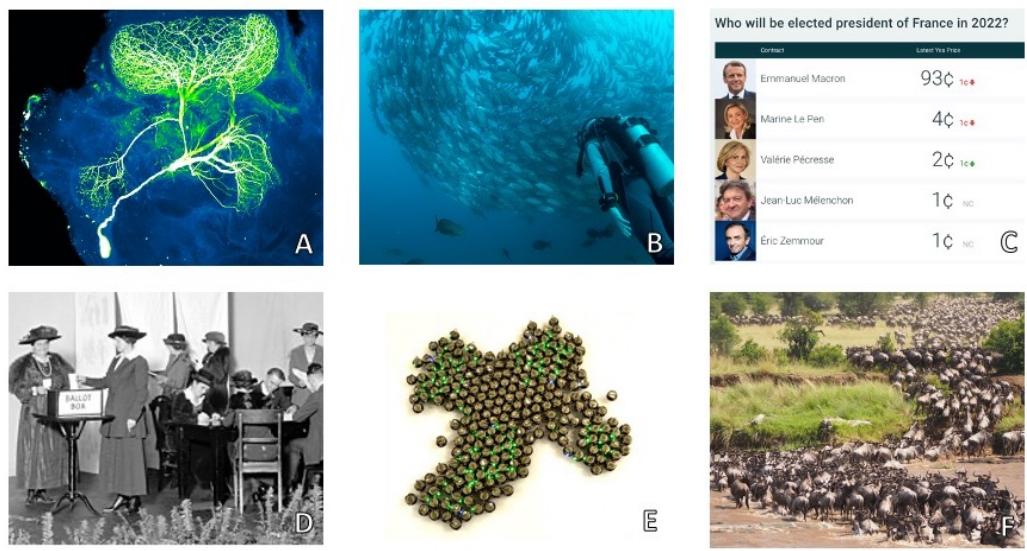

**Figure 2.** *Examples of collective intelligence systems, which process and aggregate distributed information. (A) Neurons in an insect brain (NICHD, 2015). (B) Fish school respond to the presence of a diver (van de Vendel, 2017). (C) A prediction market for who will be elected president of France in 2022 (Predictit, 2022). (D) Elections. Picture shows women voting in the United States in 1920 (Voting, 1920). (E) A swarm of robots (Carrillo-Zapata, 2019). (F) Great wildebeest migration at Serengeti National Park, Tanzania (Tung, 2019).*

With many rigorous research efforts in the past two decades in Collective Intelligence, researchers have found many factors that help or hurt the performance of human groups in solving problems and making collective decisions, predictions, or estimations. While research does not yet offer recipes for constructing a good human collective, as collective performance is the outcome of complex interactions of many variables. These findings point to a few general directions for improving the collective performance of human groups, and we offer a brief and non-exhaustive summary of relevant findings below.

**2.1 Summary of Human Collective Intelligence Findings**

**Presence of Collective Intelligence.** Similar to individual general intelligence, that one individual can excel at a wide range of tasks, such as math and music, a wide range of experiments found a similar property for human groups (Wooley et al., 2010; Wooley et al., 2015; Riedl et al., 2021), referred to as Collective Intelligence. When asking small groups to perform a wide range of tasks, including creative brainstorming, negotiation, verbal, mathematical, and moral reasoning, some groups outperform others on a wide variety of tasks. Factor analysis shows a single dominant factor explaining 43% of the variance in performance, similar to the explanatory power of individual intelligence on individual tasks (Wooley et al., 2015). These findings suggest the equivalent of general intelligence for groups.

**Group social processes more important than individual skills**. Researchers have dedicated extensive work into what makes some groups more effective than others. The consensus is that the group's social interaction processes are more important than the skill of individual members (Riedl et al., 2021). Research has found that the group member's social perceptiveness, the ability of individuals to identify social cues, improves group social processes and, consequently, group performance. This is manifested in group behaviors such as even conversational turn-taking—each group member speaks roughly equal amounts. Consequently, groups with a higher proportion of women tend to have higher collective intelligence, as being female is correlated with greater social perceptiveness. The IQ of individual members plays a much less, and some argue, negligible role.

**Diversity.** Besides the social perceptiveness of group members, research finds benefit in having a diverse group of individuals. The diversity referred to here is diversity in knowledge and cognitive models. Abundant research has found, in lab experiments, theoretical and computational models, and real-world problem-solving scenarios, the phenomenon of a "diversity bonus"—that a diverse group performs better than a homogenous group (Page, 2019; Aminpour et al., 2021). Some even find a diverse group of non-experts can outperform a homogeneous group of experts (Hong & Page, 2004). For a group of diverse agents to work together, an important aspect is cognitive alignment, such as commitment to group goals and shared beliefs (Krafft, 2019). Another line of work finds that maintaining diversity in the group is hard—conformity and traditional market forces are against it. Mann & Helbing (2016) propose alternative incentives to reward accurate minority predictions for maintaining diversity in the group—especially rewarding the minority which is right when the majority is wrong.

**Committed minorities.** There has been much evidence that the presence of committed minorities, a small fraction of individuals who are little affected by the opinions of others, can lead to substantial changes in the collective behavior of a group. Most notably, Centola et al. (2018) find through human subject experiments that a critical mass of committed individuals (around 25% in their experiments) can tip social conventions towards the direction of these committed individuals. This critical transition is also predicted by an abundance of theoretical studies (such as Xie et al., 2011). It would require further investigation to understand the precise critical mass needed in different scenarios; however, the powerful effects of committed minorities on groups can be a fruitful direction in thinking about how to elicit (or prevent) social change.

**Social influence.** An area under debate is whether and how individuals should communicate with each other in a collective. Experimental studies have found that social influence can lead to both positive and negative effects on collective performance (see Jayles et al., 2017 for an example of positive effect, and Lorenz et al., 2011 for an example of negative effect). On the one hand, letting individuals exchange information can lead to the loss of independent information, and worse collective performance (also referred to as groupthink). On the other hand, communication may help individuals deliberate and discover better answers. The conflicting findings are likely an outcome of the effect of social influence depending on several other variables. A study predicts it depends on the proportion of individuals using social information for their decisions, and whether committed minorities are present (Yang et al., 2021). Others find they also depend on the social network structure and adaptability (Almaatouq et al., 2020; Becker et al., 2017). The bottom line is that most people blindly following others does not lead to good outcomes in most scenarios.

**Better sensors.** One important component for improving group performance is to let individuals gather better information. This is especially important in forecasting and prediction tasks, such as election forecasts. This can be achieved by asking better survey questions. For example, in a method called "surprisingly popular" (Prelec et al., 2017), instead of asking people what they think, researchers also ask people what they think the majority thinks. This helps uncover surprising outcomes whose signals are suppressed by social norms, such as Trump winning the 2016 US election. Another example is to use individuals' social circles as better sensors (Galesic et al., 2021)—instead of asking individuals whom they will vote for, ask whom their friends will vote for.

The literature in this field is vast, and here we highlight a small subset of findings relevant to the performance of human groups. One theme that appears in these scientific studies is that CI requires certain conditions to appear---individuals' skills matter less than how the individuals interact, and individuals being highly correlated tend to hurt group performance.

Most CI research on human groups, such as those summarized above, focuses on improving estimation accuracies, the effectiveness of teamwork, and collective decision speed. Many aspects of these investigations have implications for GCR mitigation. Below we outline examples of current GCR mitigation efforts and discuss how CI research can be applied to mitigating GCR.

### 3. Collective Intelligence for Global Catastrophic Risk Mitigation

CI has been increasingly considered a useful part of crisis response (Vivacqua & Borges, 2010; Büscher & Thomas, 2014), and can be useful in GCR mitigation. Mitigation of GCR can naturally be divided into prevention (prevent risks from occurring), response (react before or during the hazard to limit the damage), and resilience (survive and rebuild), with a key role of adaptive collective responses to implement these mitigation measures (Cotton-Barratt et al., 2020). CI can be applied in each of these layers, in particular by enhancing predictive and adaptive ability (See Fig. 3 for a conceptual illustration of the relationships among these concepts).

### 3.1 Prediction

For prevention, identification of a possible risk and action to reduce its probability (or at least severity) are needed. The identification step is where committed minorities have historically played a clear role, often starting in the scientific community. For example, Clair Cameron Patterson built up a network concerned about the spread of lead in the environment, and the atmosphere scientists of the Crutzen-Rowland-Molina group discovered and documented ozone depletion. Their persistent research, documentation, and outreach provided the signal needed for scientific societies and later international organizations to react to the risk. Currently, the planetary defense community working on the impact risks of asteroids and the AI safety community are examples of committed minorities having the specialized concern needed to determine the risk and propose preparedness actions. These are examples of how the scientific and policy community can act with CI, by having specialized networks examining potential threats and—if the evidence becomes compelling—amplifying their signal through a social influence mechanism.

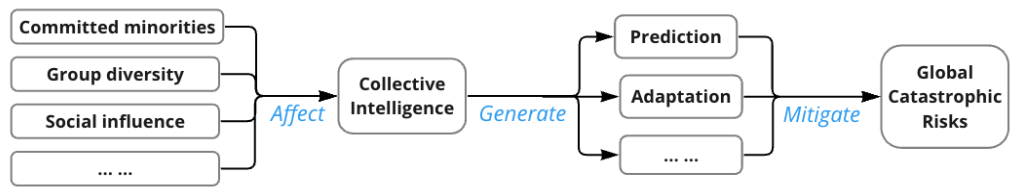

*Figure 3.* A conceptual illustration of the concepts in this article and their connections. Collective intelligence as infrastructure for mitigating global catastrophic risks through avenues such as prediction and adaptation.

Predicting GCRs ahead of time is challenging because they represent unprecedented problems, or occur in domains with very little data. There has been significant interest in estimating the likelihood of different GCRs, especially due to the need for prioritization and making trade-offs among risks. This has been done using a variety of methods that aggregate expert or public opinion (Beard et al., 2020). At the simplest, these methods consist of taking the median estimate, while more sophisticated approaches give weightings based on consistency (Prelec, 2004; Frank et al., 2017) or past predictive performance, involve structured deliberative processes to build a consensus, or are prediction markets where (updateable) bets are made.

One mechanism for prediction discussed at length in the CI literature is prediction markets, where participants bet on future outcomes, and the prices of bets indicate the joint belief in their likelihoods (Wolfers & Zitzewitz, 2006). Here, distributed and diverse knowledge is aggregated through a price mechanism. In theory, attempts to manipulate the price can even increase the accuracy by injecting capital exploitable by informed participants (Hanson & Opera, 2009). Existing prediction markets contain bets on various natural or anthropogenic disasters. In an experiment, a prediction market was applied to forecasting epidemic disease rates and performed more accurately than predictions from extrapolating historical trends (Li et al., 2016). However, prediction markets have so far been limited by anti-gambling laws in many countries, lack of liquidity/participants, and the time preferences of participants.

Similar to research on prediction markets, work on forecasting tournaments, which is not limited by anti-gambling laws, has produced methodologies for eliciting short- and medium-term forecasts that are consistently better than individual expert forecasts, demonstrating CI. These methodologies involve weighted aggregation of many individual estimates, identifying top performers and grouping them together into teams, and supporting learning, updating, and debiasing (Tetlock & Gardner, 2016; Tetlock et al., 2017). Inspired by the success of forecasting tournaments and prediction markets, metaculus.com is a community for group forecasting using many of the same aggregation mechanisms, plus others like gamified scoring and reputation to encourage higher accuracy. The site covers a vast variety of real-world events, including disasters all the way up to existential risk. The site also did well compared to both the public and experts in forecasting during the COVID-19 pandemic (Recchia et al., 2021).

Often preparation, rapid detection, and response are more feasible than predicting the precise moment a GCR will happen. Rapid detection can draw on the distributed and sensitive nature of CI to find early signals of a hazard, amplify the warning, and trigger pre-planned responses. One experiment demonstrated how rapid social mobilization using social media and an incentive mechanism to solve an "impossible" search problem (finding ten weather balloons located across the US) was able to succeed within 9 hours. Here the winning team CI was enabled by an incentive mechanism that rewarded not just direct search but recruiting likely candidates, plus information sharing across the participants. Other teams recruited suitable pre-existing communities (Pickard et al., 2011). Similar experiments have demonstrated fast finding of people (<12 hours) globally (Rutherford et al., 2013). These results suggest that rapid, CI-mediated crisis response such as finding pathogen carriers or dangerous objects is possible.

### 3.2 Adaptation

After the initial forecast, adaptivity may become essential. Decision-making and governance relevant to global catastrophic risk are complex, and naturally take place inside a collective cognitive system that is part of the greater system: complexity, reflexivity, and uncertainty cannot be avoided (Fisher & Sandberg, 2022). Responding to GCR requires adaptive governance. In evolving disaster situations, flexible solutions and the ability to innovate on the fly can be crucial. The responses need to be more adaptive than the hazard, which typically implies a great need for generating diverse solutions, rapid spread of successful solutions, and forming tight feedback loops to help evaluate and improve the solutions. CI has a clear advantage here over centralized control systems that tend to act slowly and with less information. Such top-down approaches may be more suited for pre- or post-disaster preparation and systemic recovery where timescales are long, optimized global solutions perhaps possible and desirable, and especially when a severe GCR needs to be avoided at all costs. Adaptive governance has demonstrated success in managing common-pool resources in a decentralized fashion (Ostrom, 1990). Safety can be thought of as a special common pool resource that can be managed in this way. GCR raises the stakes by making the threat global and hence requires global adaptive governance, a challenge for CI.

One component that can improve adaptive governance in a GCR situation is effective data-sharing, a lesson learned especially from the COVID-19 pandemic. Data aggregation sites such as ourworldindata.org played a key role in providing various agencies and networks with vital information while they struggled to get data from governments in useful formats. To enable effective CI, constructing better data-sharing infrastructure may be necessary.

Another component that can improve adaptive governance is resiliency. Resiliency can be viewed as the capacity of a system exposed to a shock or stress to adapt and survive by changing its non-essential attributes and rebuilding itself (Downes, 2013). Resiliency, both during and after a catastrophe, is often aided by collective memory: adaptations to past disasters remain distributed among community members (and sometimes, stigmergically (Marsh & Onof, 2008), in the environment like Japanese tsunami stones reminding of long-term danger). Collective memory may have a limited range (Fanta et al., 2019) but can both exceed individual lifespans and allow distributed reactions. It also interacts with other collective attitudes such as risk-taking and trust (Viglione et al., 2014). Another form of collective risk cognition affecting resiliency is via insurance pricing, where insurers set premiums based on aggregated and estimated risk, creating incentives. For example, in response to tsunami risks, insurance incentives can incentivize homeowners and builders to build where the risks are reduced.

Many forms of adaptations to risk in human society take the form of institutions, whether formal (such as fire services, disaster agencies, the Intergovernmental Panel on Climate Change) or informal (such as cautionary tales, safety practices). They can be seen as integrated systems of rules that structure social interactions (Hodgson, 2015): hence a way of structuring or creating CI that (when done well) promotes resilience. This institution-forming often occurs in a reactive way when disaster strikes, which may not be acceptable for the most serious risks.

CI is often described as spontaneously emergent, but this is rare. Usually, it is enabled by the existence of certain structures—suitable abilities, platforms, and incentives that make forming a shared cognitive system easier and allow it to have strong effects. The CI literature suggests such structures need to give preference to cognitively diverse and socially attuned individuals, and egalitarian group social processes. Individual influences need to be homogeneously distributed, and there is not too much correlation in the system, such as a few individuals having an overwhelming influence on the group. Both sets of ideal properties generating CI are counter to elements in the current Western society. In companies, individual employees are often evaluated based on individual-level output, rather than how well they facilitate teamwork. In online social networks, a small number of actors can have an overwhelming influence on the overall network. The CI literature suggests these conditions could hinder collective intelligence. Thus, there is a need to use CI research to design and improve these structures—such as the types of institutions and market incentives—for adaptive governance.

## 4. Discussion

Systems can be stable by resisting shocks directly ("toughness"), by returning to their equilibria through internal feedback (homeostasis, such as how mammals regulate body temperature), by dynamic stability (constant nudging the system to stay in an intrinsically unstable equilibrium, for example, riding a bicycle), to adaptive stability (the system changes structure to respond to the shocks, for example, buildings add safety stairs and smoke alarms to improve resilience to fires). In human systems facing disaster, sufficient toughness can rarely be achieved but feedback and adaptation are both possible and common. CI represents the distributed, bottom-up approach to achieving these without central control, which can either work on its own or in combination with top-down governance.

A central issue in GCR mitigation is how to reconcile individuals with conflicting goals and interests. This is especially manifested in concerns derived in game-theoretic analyses, such as prisoners' dilemma and tragedy of the commons. Collective intelligence offers a broad and, we think, hopeful, perspective on this issue. These differences may first appear as an obstacle to coordination, while in the collective intelligence perspective, these differences, if harnessed with the right aggregation mechanism, are a source of strength. Take the example of a flock of fish. Each fish senses their local environment—food, temperature, potential predator, etc., and has a different preferred direction of where to go next. Nevertheless, the collective effectively aggregate information from all fish through local interactions in movement (Lopez et al., 2012). The collective's movement responds to a much wider environment, more than the range of any individual fish. Thus, collective intelligence can be generated from different individuals acting on different information and wanting different outcomes. The differences should be seen as a resource, and the key question should be how to harness and aggregate the individuals' differences in productive ways.

It can be useful to take advantage of the transdisciplinary nature of CI and learn from other disciplines. Biological systems, through evolution, have devised efficient solutions to complex CI problems. Besides the example of fish, each neuron in the human brain receives different, and sometimes conflicting information, but the brain harnesses the inconsistent signals into a cohesive understanding of the world. What systems in nature face similar challenges in processing distributed information as we do in collective action and decision-making? Following the biomimicry concept in engineering, social system researchers could gain valuable lessons from biological systems' approaches to CI, informing solutions to societal challenges.

## 5. Conclusions

The emerging field of GCR mitigation presents significant challenges with substantial safety implications. Insights from CI research can provide valuable guidelines to aid policymakers in their decision-making processes, particularly in incorporating diverse and socially-aware groups and prioritizing egalitarian interactions. CI research highlights the role of social processes within a group's capacity for collective decision-making. Future studies should examine how shifts in social interaction and media distribution affect large-scale decisions, such as those made in democratic elections. While most CI studies occur in routine contexts, such as brain-storming or medical diagnosis, future research should examine whether the factors improving group performance in these situations also apply to high-stake, urgent situations like addressing GCR mitigation, affecting a vast population.

CI research also suggests new strategies for GCR mitigation, like leveraging committed minorities for prediction and utilizing collective memory for adaptation. Beyond these areas, one promising direction of future research is collective intelligence in human-AI groups, where the different kinds of cognition may complement each other (Guszcza et al., 2022). AI methods can obviously outsource some human cognitive tasks, but could also perform tasks not suited to human thinking (e.g., watching for high dimensional patterns). However, how to design human-AI collaborations well remains under-explored. Another important topic worth further analysis is how the learnings from CI research can be used in the existing disaster management frameworks such as the Sendai framework and the EU Civil Protection Mechanism. In many ways, the frameworks represent a form of institution-based CI already applied to risk management, but it is plausible that other CI methods may help improve their performance if they can be integrated with existing structures.

We encourage CI researchers to think more about using CI in practice, especially in GCR scenarios representing perhaps the most important societal challenges. We encourage GCR researchers to expand their approach, which is centered on studying the physical sciences behind each risk scenario, to further engage with the behavioral sciences, especially the study of collective intelligence, to formulate useful theoretical frameworks for the reduction of risks across domains. We encourage general behavioral researchers to become aware and take advantage of the transformative opportunities of making an impact on the most important societal challenges through engaging with these two transdisciplinary efforts.